\documentclass[preprint, 3p, sort]{elsarticle}
\usepackage{dcolumn}
\usepackage{booktabs,tabularx} 
\usepackage{booktabs}
\usepackage{textcomp}
\usepackage{amssymb}
\usepackage{amsthm}
\usepackage[mathlines]{lineno}
\usepackage{graphicx}
\usepackage{units}
\usepackage{url}
\usepackage{amsmath}
\usepackage{amsfonts}
\usepackage{bm}
\usepackage{textcomp}
\usepackage{subfigure}
\usepackage{multicol}
\usepackage{multirow}
\usepackage{verbatim}
\usepackage{rotating}
\usepackage[colorlinks,linkcolor=black,citecolor=red]{hyperref}
\usepackage{float}
\usepackage{epsfig}
\usepackage{dcolumn}
\usepackage{bm}
\usepackage{color}
\usepackage{pstricks}
\usepackage{pst-node}
\usepackage{times}
\usepackage{indentfirst}
\usepackage[english]{babel}
\usepackage{epstopdf}
\addto{\captionsenglish}{%

}
\journal{Physics Letters B}


\usepackage[font=footnotesize]{caption}

\usepackage{overpic}

\newcommand{\mevcc}{\,\mbox{MeV}/c^2}

\newcommand{\gev}{\,\mbox{GeV}}
\newcommand{\gevc}{\,\mbox{GeV}/c}
\newcommand{\gevcc}{\,\mbox{GeV}/c^2}

\newcommand{\jpsi}{J/\psi}
\newcommand{\ra}{\rightarrow}
\newcommand{\inv}{+{\rm invisible}}

\newcommand{\xim}{\Xi^-}
\newcommand{\xip}{\bar{\Xi}^+}
\newcommand{\piz}{\pi^0}
\newcommand{\pip}{\pi^+}
\newcommand{\pim}{\pi^-}
\newcommand{\pb}{\bar{p}}
\newcommand{\nb}{\bar{n}}
\newcommand{\emc}{E_{\rm EMC}}

\newcommand{\lmd}{\Lambda}
\newcommand{\lmdb}{\bar{\Lambda}}

\biboptions{numbers,sort&compress}
\begin{document}
\begin{frontmatter}
\title{{\boldmath \bf Search for a dark baryon in the $\xim\ra\pim\inv$ decay}}
\author{\begin{small}
\begin{center}
M.~Ablikim$^{1}$, M.~N.~Achasov$^{4,c}$, P.~Adlarson$^{77}$, X.~C.~Ai$^{82}$, R.~Aliberti$^{36}$, A.~Amoroso$^{76A,76C}$, Q.~An$^{73,59,a}$, Y.~Bai$^{58}$, O.~Bakina$^{37}$, Y.~Ban$^{47,h}$
% ...
\\
\vspace{0.2cm}
(BESIII Collaboration)\\
\vspace{0.2cm} {\it
$^{1}$ Institute of High Energy Physics, Beijing 100049, People's Republic of China\\
% ...
$^{4}$ Budker Institute of Nuclear Physics SB RAS (BINP), Novosibirsk 630090, Russia\\
% ...
$^{36}$ Johannes Gutenberg University of Mainz, Johann-Joachim-Becher-Weg 45, D-55099 Mainz, Germany\\
$^{37}$ Joint Institute for Nuclear Research, 141980 Dubna, Moscow region, Russia\\
% ...
$^{47}$ Peking University, Beijing 100871, People's Republic of China\\
% ...
$^{58}$ Southeast University, Nanjing 211100, People's Republic of China\\
$^{59}$ State Key Laboratory of Particle Detection and Electronics, Beijing 100049, Hefei 230026, People's Republic of China\\
% ...
$^{73}$ University of Science and Technology of China, Hefei 230026, People's Republic of China\\
% ...
$^{76}$ University of Turin and INFN, (A)University of Turin, I-10125, Turin, Italy; (B)University of Eastern Piedmont, I-15121, Alessandria, Italy; (C)INFN, I-10125, Turin, Italy\\
$^{77}$ Uppsala University, Box 516, SE-75120 Uppsala, Sweden\\
% ...
$^{82}$ Zhengzhou University, Zhengzhou 450001, People's Republic of China\\

\vspace{0.2cm}
$^{a}$ Deceased\\
% ...
$^{c}$ Also at the Novosibirsk State University, Novosibirsk, 630090, Russia\\
% ...
$^{h}$ Also at State Key Laboratory of Nuclear Physics and Technology, Peking University, Beijing 100871, People's Republic of China\\
% ...

}
\end{center}
\end{small}
}

\date{\today}

\begin{abstract}
A search for a dark baryon is performed for the first time in the two-body decay $\xim\ra \pim\inv$ using $(10.087\pm0.044)\times10^{9}$ $\jpsi$ events collected at a center-of-mass energy of $\sqrt{s}=3.097\gev$ with the BESIII detector at the BEPCII collider. No significant signal is observed, and the 90\% (95\%) confidence level upper limits on the branching fraction $\mathcal{B}(\xim\ra\pim\inv)$ are determined to be $4.2\times10^{-5}$ ($5.2\times10^{-5}$), $6.9\times10^{-5}$ ($8.4\times10^{-5}$), $6.5\times10^{-4}$ ($7.6\times10^{-4}$), $1.1\times10^{-4}$ ($1.3\times10^{-4}$) and $4.5\times10^{-5}$ ($5.5\times10^{-5}$), under the dark baryon mass hypotheses of 1.07$\gevcc$, 1.10$\gevcc$, $m_\lmd$ (1.116$\gevcc$), 1.13$\gevcc$, and 1.16$\gevcc$, respectively. The constraints obtained on the Wilson coefficients $C_{u s, s}^L$ and $C_{u s, s}^R$ are more stringent than the previous limits derived from the LHC searches for the colored mediators.

\end{abstract}
\end{frontmatter}

\begin{multicols}{2}
\section{Introduction}

The existence of dark matter (DM) is strongly supported by astrophysical and cosmological observations, yet its nature remains one of the unsolved problems within the Standard Model (SM) of particle physics. 
One hint to its identity is the similarity between the DM and baryon densities, $\rho_{\rm{DM}} \approx 5.4\,\rho_{\rm{baryon}}$~\cite{density}, suggesting a potential connection between their origins and motivating the existence of dark sector particles charged under a baryon gauge symmetry with masses at the $\gevcc$ scale~\cite{dark1, dark2}. 
The baryonic dark sector has been further motivated by a long-standing discrepancy between the neutron lifetime measured in beam and bottle experiments, which could be resolved if the neutron decays into dark states carrying baryon number with a branching fraction (BF) at the level of $1\%$~\cite{neutron}.
Furthermore, provided that $B$ mesons decay into dark sector antibaryons with a BF larger than 0.01\%~\cite{BMeson1}, the $B$-Mesogenesis mechanism~\cite{BMeson2, BMeson3} can explain the asymmetry between visible matter and antimatter and also the origin and nature of dark matter. 

The dark sector anti-baryon has been searched for in decays of $B$ mesons by the \textit{BABAR} experiment~\cite{BABAR1, BABAR2}. 
Complementary to this, hyperons offer the opportunity to search for the baryonic dark sector through decays into final states containing dark baryons~\cite{factory}, which appear as missing energy in a detector. Invisible decays of the $\lmd$ baryon have been searched for by the BESIII experiment~\cite{lmdinv}, constraining the BF of the $\lmd\ra\rm{invisible}$ decay to be less than $7.4\times10^{-5}$ at the 90\% confidence level (C.L.). 
Exploring pionic hyperon decays with invisible signatures offers access to a broader range of dark baryon masses and an improved experimental sensitivity~\cite{dark_baryons}. The BF of the pionic decay of the $\xim$ baryon with an invisible signature may be as large as $10^{-3}$ and is not constrained by the SN 1987A cooling bound~\cite{dark_baryons}, thus rendering it a priority target for laboratory searches.

This Letter reports a search for a dark baryon in the two-body decay $\xim\ra \pim\inv$, where the $\xim$ candidate is identified by tagging a $\xip$ decaying to $\pip\lmdb(\ra\pb\pip)$ on its recoiling side~\cite{tag}.
The analysis exploits around $10^7$ $\xim\xip$ hyperon pairs produced from $(10.087\pm0.044)\times10^{9}$ $\jpsi$ decays~\cite{jpsinum} collected at a center-of-mass (CM) energy of $\sqrt{s}=3.097\gev$ with the BESIII detector at the BEPCII collider.
The charge-conjugated decay $\xip \to \pip \inv$ is not investigated in this analysis due to the dominant background from $\xip \to\pip \lmdb (\to \nb \piz)$, where the interactions of antineutrons with the detector material produce widespread showers, complicating accurate simulations and the determination of a clean control sample for corrections.
A semi-blind procedure is performed to avoid possible bias, where
approximately 10\% of the data sample is used to validate the
analysis procedures  before performing the final analysis on the full
data set.

\section{BESIII detector and Monte Carlo simulation}

The BESIII detector~\cite{Ablikim:2009aa} records symmetric $e^+e^-$ collisions 
provided by the BEPCII storage ring~\cite{Yu:IPAC2016-TUYA01} in the CM energy range from 1.84 to 4.95~GeV. The BESIII detector has collected large data samples in this energy region~\cite{Ablikim:2019hff}. The cylindrical core of the BESIII detector covers 93\% of the full solid angle and consists of a helium-based multilayer drift chamber~(MDC), a plastic scintillator time-of-flight system~(TOF), and a CsI(Tl) electromagnetic calorimeter~(EMC), which are all enclosed in a superconducting solenoidal magnet providing a 1.0~T magnetic field. The magnetic field was 0.9~T in 2012, which affects 11\% of the total $J/\psi$ data. The solenoid is supported by an
octagonal flux-return yoke with resistive plate counter muon identification modules interleaved with steel. The charged-particle momentum resolution at $1~{\rm GeV}/c$ is $0.5\%$, and the ${\rm d}E/{\rm d}x$ resolution is $6\%$ for electrons from Bhabha scattering. The EMC measures photon energies with a resolution of $2.5\%$ ($5\%$) at $1$~GeV in the barrel (end-cap) region. The time resolution in the TOF barrel region is 68~ps, while that in the end-cap region is 110~ps. The end-cap TOF
system was upgraded in 2015 using multigap resistive plate chamber technology, providing a time resolution of 60~ps, which benefits 87\% of the data used in this analysis~\cite{etof1,etof2,etof3}.

Simulation samples produced with a {\sc geant4}-based~\cite{geant4} Monte Carlo (MC) package, which includes the geometric description~\cite{geometric,li2024,liu2009,you2008} of the BESIII detector and the detector response, are used to determine detection efficiencies
and to estimate backgrounds. The simulation models the beam-energy spread and initial-state radiation in the $e^+e^-$ annihilations with the generator {\sc
kkmc}~\cite{ref:kkmc}. The inclusive MC sample includes the production of the $J/\psi$ resonance incorporated in {\sc kkmc}. All particle decays are modeled with {\sc evtgen}~\cite{ref:evtgen} using BFs 
either taken from the Particle Data Group~\cite{pdg}, when available, or otherwise estimated with {\sc lundcharm}~\cite{lundcharm1, lundcharm2}. Final-state radiation from charged final-state particles is incorporated using the {\sc photos} package~\cite{photos}. To study the tagging efficiency of the $\xip\ra\pip\lmdb(\ra\pb\pip)$ decay, the MC sample of $\jpsi\ra\xim(\ra {\rm anything})\xip(\ra\pip\lmdb(\ra\pb\pip))$ is generated according to its helicity decay amplitudes as detailed in Refs.~\cite{Xi0_para, Xim_para, Lmd_para}.
The signal decay $\jpsi\ra\xim(\ra\pim\chi)\xip(\ra\pip\lmdb(\ra\pb\pip))$ is generated according to its helicity decay amplitudes, where $\chi$ designates a dark baryon with an invisible signature and the decay-asymmetry parameter of $\xim\ra \pim\chi$ is assumed to be the same as that of the $\xim\ra\pim\lmd$ decay~\cite{Xim_para}. To satisfy the kinematic constraints while accounting for the background conditions, the signal events are generated under dark baryon mass ($m_\chi$) hypotheses of 1.07$\gevcc$, 1.10$\gevcc$, $m_\lmd$, 1.13$\gevcc$, and 1.16$\gevcc$, where the $m_\lmd$ is the known mass of the $\lmd$~\cite{pdg}.

\section{Data analysis}
\subsection{Analysis method}

For the signal decay $\xim\ra\pim\inv$, the $\xim$ hyperon is inferred by reconstructing the $\xip$ decay in the events of $\jpsi\ra\xim\xip$ at $\sqrt{s}=3.097\gev$. 
The $\xip$ candidates, which constitute the single-tag (ST) sample, are reconstructed with the dominant decay $\xip\ra\pip\lmdb(\ra\pb\pip)$. Then the double-tag (DT) event is formed by reconstructing the signal decay $\xim\ra\pim\inv$ in the system recoiling against the reconstructed $\xip$ hyperon. The absolute BF of the signal decay is determined by
\begin{equation}
\mathcal{B}(\xim\ra\pim\inv)=\frac{N_{\mathrm{DT}}^{\rm{obs}} / \epsilon_{\mathrm{DT}}}{N_{\mathrm{ST}}^{\rm{obs}} / \epsilon_{\mathrm{ST}}},
\end{equation}
where $N_{\rm{ST}}^{\rm{obs}}$ $(N_{\rm{DT}}^{\rm{obs}})$  is the observed ST (DT) yield and $\epsilon_{\mathrm{ST}}$ $\left(\epsilon_{\mathrm{DT}}\right)$ is the corresponding detection efficiency.

\subsection{ST selection and yields}

Charged tracks detected in the MDC are required to be within a polar angle ($\theta$) range of $|\rm{cos\theta}|<0.93$, where $\theta$ is defined with respect to the $z$-axis, which is the symmetry axis of the MDC. Particle identification~(PID) for charged tracks combines measurements of the specific ionization energy loss (${\rm d}E/{\rm d}x$) in the MDC and the flight time in the TOF to form likelihoods $\mathcal{L}(h)~(h=p,K,\pi)$ for each hadron $h$ hypothesis. Tracks are identified as protons (pions) when the proton (pion) hypothesis has the greatest likelihood among these three hypotheses.

To reconstruct $\lmdb$ and $\xip$ candidates, vertex fits~\cite{vertex} are applied to the $\pb\pip$ and the $\lmdb\pip$ combinations, respectively, given that $\lmdb$ and $\xip$ have relatively long lifetimes. 
A vertex fit is performed to obtain the decay vertex of the $\lmdb$ ($\xip$). The production vertices of the $\lmdb$ and $\xip$ are the decay vertex of the $\xip$ and the $e^+e^-$interaction point, respectively.
A subsequent vertex fit is performed using the parameters of the production vertex, decay vertex, and the $\lmdb$ ($\xip$) flight direction. To suppress background from non-$\lmdb$ (non-$\xip$) processes, the decay length of the $\lmdb$ ($\xip$) is required to be larger than zero. The decay length is the distance from the production vertex to the decay vertex, and negative decay lengths can be caused by the detector resolution. 
The $\pb\pip$ and $\lmdb\pip$ combinations are chosen with the minimum value of the sum of $|M_{\pb\pip}-M_{\lmdb}|$ and $|M_{\lmdb\pip}-M_{\xip}|$, and $|M_{\pb\pip}-M_{\lmdb}|$ is required to be less than $4\mevcc$, where $M_{\lmdb}$ ($M_{\xip}$) is the known mass of the $\lmdb$ ($\xip$)~\cite{pdg}.
The recoiling mass against the reconstructed $\xip$ candidate is defined as
\begin{equation}
M_{\bar{\Lambda} \pi^{+}}^{\rm recoil}=\sqrt{\left(E_{\rm CM}-E_{\bar{\Lambda} \pi^{+}}\right)^2/c^{4}-\vec{P}_{\bar{\Lambda} \pi^{+}}^2/c^{2}},
\end{equation}
where $E_{\rm CM}$ is the CM energy, and $E_{\bar{\Lambda} \pi^{+}}$ and $\vec{P}_{\bar{\Lambda} \pi^{+}}$ are the energy and momentum of the selected $\lmdb\pip$ system defined in the CM system, which have been corrected during the vertex fits. To further suppress backgrounds, the $M_{\bar{\Lambda} \pi^{+}}^{\rm recoil}$ is required to be in the $\Xi^{-}$ signal region, defined as (1.290, 1.345)$\gevcc$, corresponding to the region of approximately three times the resolution around the signal peak.

\begin{figure}[H]
    \begin{center}
        \begin{overpic}[width=7.5cm]{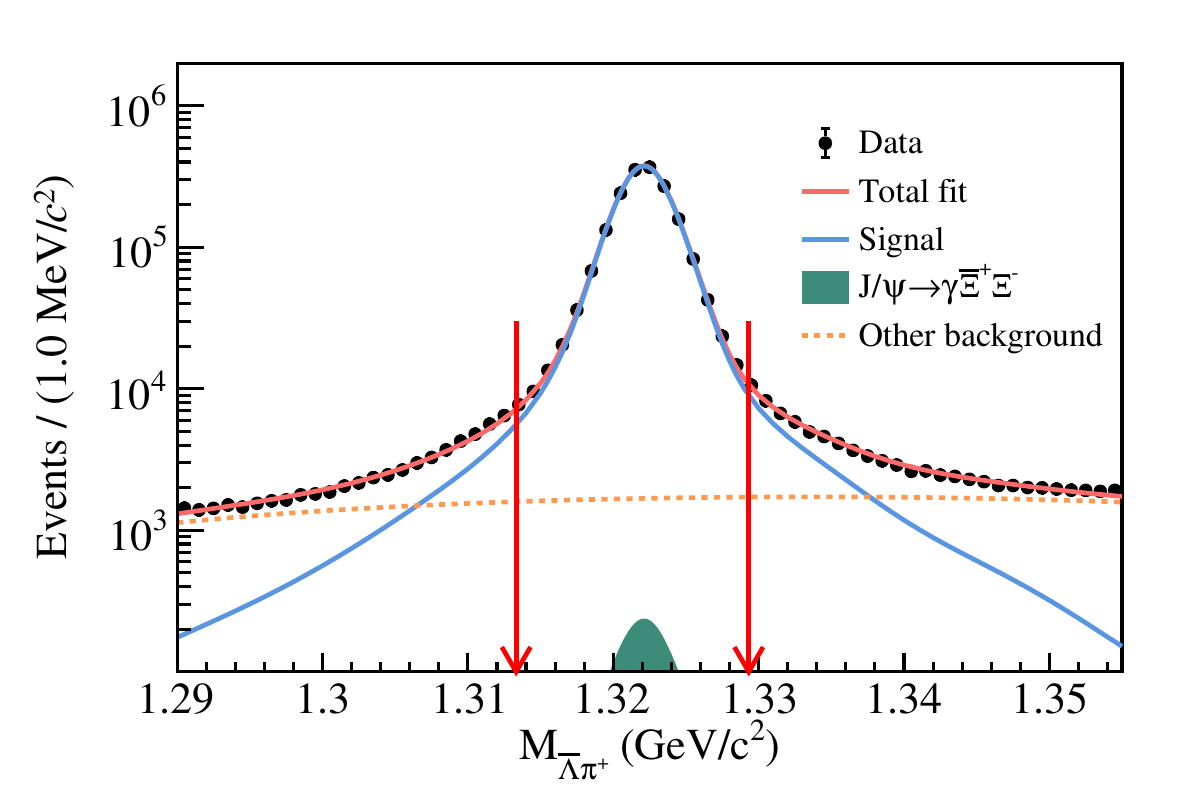}
        \end{overpic} 
    \caption{The $M_{\lmdb\pip}$ distribution of ST candidates. The black dots with error bars are data and the red solid curve is the fit result. The $\jpsi\ra\gamma\xip\xim$ peaking background, signal and other nonpeaking background are represented by the green shaded region, blue solid and orange dashed curves, respectively.
    The red arrows indicate the ST signal window.}
        \label{fig:STfit}
    \end{center}
\end{figure}

A binned maximum likelihood fit is performed to the $M_{\lmdb\pip}$ distribution
to obtain the ST yield. In the fit, the signal shape is modeled by the MC-simulated shape convolved with a Gaussian function to account for the resolution difference between data and MC simulation. The signal region is defined as $|M_{\lmdb\pip}-M_{\xip}|<8\mevcc$.
By analyzing the inclusive MC sample with the help of a generic event type
analysis tool, TopoAna~\cite{topoana}, the peaking background is mainly from $\jpsi\ra\gamma\xip\xim$ and is estimated to contribute $1199\pm 35$ events in the signal region. Other nonpeaking background is described by a second-order Chebyshev polynomial function. The fit result is shown in Fig.~\ref{fig:STfit} and 
the ST yield extracted from the fit is $(1813.4\pm1.4)\times10^{3}$ after subtracting the contribution from the $\jpsi\ra\gamma\xip\xim$ background.
The ST detection efficiency is evaluated using the signal MC
sample and determined to be $(26.03\pm0.03)\%$, where the uncertainty is statistical only. The BF of $\jpsi\ra\xim\xip$ is calculated according to the observed ST yield and the corresponding ST efficiency and is found to be compatible with the previous BESIII measurement~\cite{br_xi} within uncertainties.

\subsection{DT selection and yields}

The signal decay $\xim\ra \pim\inv$ is searched for using the remaining tracks recoiling against the ST $\xip$ candidates. The following criteria are applied to select the signal candidates and suppress the backgrounds mainly from the $\xim\ra\pim\lmd(\ra p\pim)$ and $\xim\ra\pim\lmd(\ra n\piz)$ processes.
Exactly one additional negatively charged particle has to be reconstructed for the DT candidate events and it must be identified as a pion. A one-constraint (1C) kinematic fit is performed under the hypothesis of $\jpsi\ra \pb\pip\pip\pim\inv$. The fit constrains the mass of the invisible particle to  $m_\chi$. The $\chi^2$ value of the kinematic fit ($\chi^2_{\pim\chi}$) must be less than 20. 
Except for the case of $m_\chi=m_\lmd$, another 1C kinematic fit is performed by constraining the mass of the missing particle to the known $\lmd$ mass. The obtained $\chi^2$ value ($\chi_{\pim\lmd}^2$) is required to be larger than $\chi_{\pim\chi}^2$. 
The four-momentum of the DT pion and $\xim$ is obtained from the 1C kinematic fit that constrains the mass of the invisible particle to $m_\chi$. The momentum of $\pim$ in the CM frame of $\xim$ ($P_{\pim}$) is required to lie within lower and upper bounds optimized by maximizing the Punzi significance~\cite{Punzi}, which is defined as $\varepsilon/(1.5+\sqrt{N_{\rm bkg}})$. Here, $\varepsilon$ denotes the signal efficiency obtained from the signal MC sample and $N_{\rm bkg}$ is the number of background events obtained from the background MC samples. The signal momentum windows are determined to be (0.184, 0.192), (0.152, 0.160), (0.132, 0.146), (0.118, 0.126), and (0.070, 0.082) $\gevc$ under the $m_\chi$ hypotheses of 1.07$\gevcc$, 1.10$\gevcc$, $m_\lmd$, 1.13$\gevcc$, and 1.16$\gevcc$, respectively.
The DT detection efficiencies are evaluated using the signal MC
samples and determined to be $(12.87\pm0.03)\%$, $(11.14\pm0.03)\%$, $(13.25\pm0.04)\%$, $(7.50\pm0.03)\%$ and $(10.28\pm0.03)\%$ under the $m_\chi$ hypotheses of 1.07$\gevcc$, 1.10$\gevcc$, $m_\lmd$, 1.13$\gevcc$, and 1.16$\gevcc$, respectively, where the uncertainties are statistical only.

After applying the discussed selection criteria, the dominant background events are from the $\xim\ra\pim\lmd(\ra n\piz)$ process, where the neutron and photon showers from $\piz$ decays would deposit energy in the EMC. Since the dark baryon has an invisible signature in the detector, the energy sum of all the showers in the EMC, $\emc$, can be utilized as a discriminator to extract the DT yield. The showers are required to be located in either the barrel region ($|\cos\theta|<0.80$) or end-cap region ($0.86<|\cos\theta|<0.92$).
The isolation angle criterion is applied to exclude showers that originate from charged tracks, where the angle subtended by the shower in the EMC and the closest charged track at the EMC must be greater than 10 degrees (20 degrees for $\pb$ tracks since anti-protons interact strongly with nuclei) as measured from the interaction point. To suppress electronic noise and showers unrelated to the event, the difference between the EMC time and the event start time is required to be within [0, 700]~ns.

For the signal events, the energy deposit mainly comes from the
interaction between the $\pb$ and the detector, under the condition
that the induced showers are already suppressed through the isolation
angle criteria. However, due to difficulties in accurately modeling
anti-proton interactions with the detector material using the {\sc
  geant4} package, the raw simulation of $\emc$ deviates from
data~\cite{pinv}. To correct this discrepancy, a control sample of
$\jpsi\ra\xim(\ra\pim\lmd(\ra p\pim))\xip(\ra\pip\lmdb(\ra\pb\pip))$
is selected and a data-driven approach is applied. The four-momenta of
the final state particles in the control sample are obtained through a
6C kinematic fit, where the invariant mass of the $\pb\pip$ ($p\pim$)
combination is constrained to the $\lmdb$ ($\lmd$) mass, and the total
four-momentum of the final-state particles is constrained to that of
the $e^+e^-$ system. The $\emc$ of signal events is randomly sampled
from the shape template obtained from the data control sample,
according to the MC-truth information of the momentum and polar angle
of the anti-proton.  The contribution from additional $p\pim$ tracks
in the control sample is already estimated and eliminated in the $\emc$ shape template using an MC sample of
$\jpsi\ra\xim(\ra\pim\lmd(\ra p\pim))\xip(\ra\pip\lmdb(\ra\pb\pip))$,
where only the $p$ and $\pim$ tracks from $\lmd$ are allowed to
interact with the detector material in the exclusive simulation. 

For the background events from $\xim\ra\pim\lmd(\ra n\piz)$, $\emc$ is divided into two parts, $\emc^{\piz}$ and $\emc^{\rm other}$. The $\emc^{\piz}$ is the energy from $\piz$ of the signal side and is obtained through the MC simulation, which describes the interactions of photons or electrons with the material with sufficient accuracy. The 
$\emc^{\rm other}$ originates from other sources from charged tracks, neutrons, and noise unrelated to the event. The shape of $\emc^{\rm other}$  is corrected using a data-driven approach based on a control sample of $\jpsi\ra\xim(\ra\pim\lmd(\ra n\piz))\xip(\ra\pip\lmdb(\ra\pb\pip))$. The four-momenta of the final state particles in the control sample are obtained through a 4C kinematic fit, where the neutron is considered as a missing track, the invariant mass of $\pb\pip$ ($n\gamma\gamma$) combination is constrained to the $\lmd$ mass, the invariant mass of $\gamma\gamma$ combination is constrained to the $\piz$ mass, and the total four-momentum of the final-state particles is constrained to that of the $e^+e^-$ system.
$\emc^{\rm other}$ is assigned with a random value from the shape template obtained from the data control sample, according to the MC-truth information of the momentum and polar angle of the anti-proton and neutron. For other background events in the inclusive MC sample, $\emc$ is corrected using the control sample where $\lmd$ decays to either $p\pim$ or $n\piz$, depending on whether a neutron is involved in the final state.

\begin{figure*}[ht] 
  \centering
  \mbox
  {
  \begin{overpic}[width=0.33\textwidth]{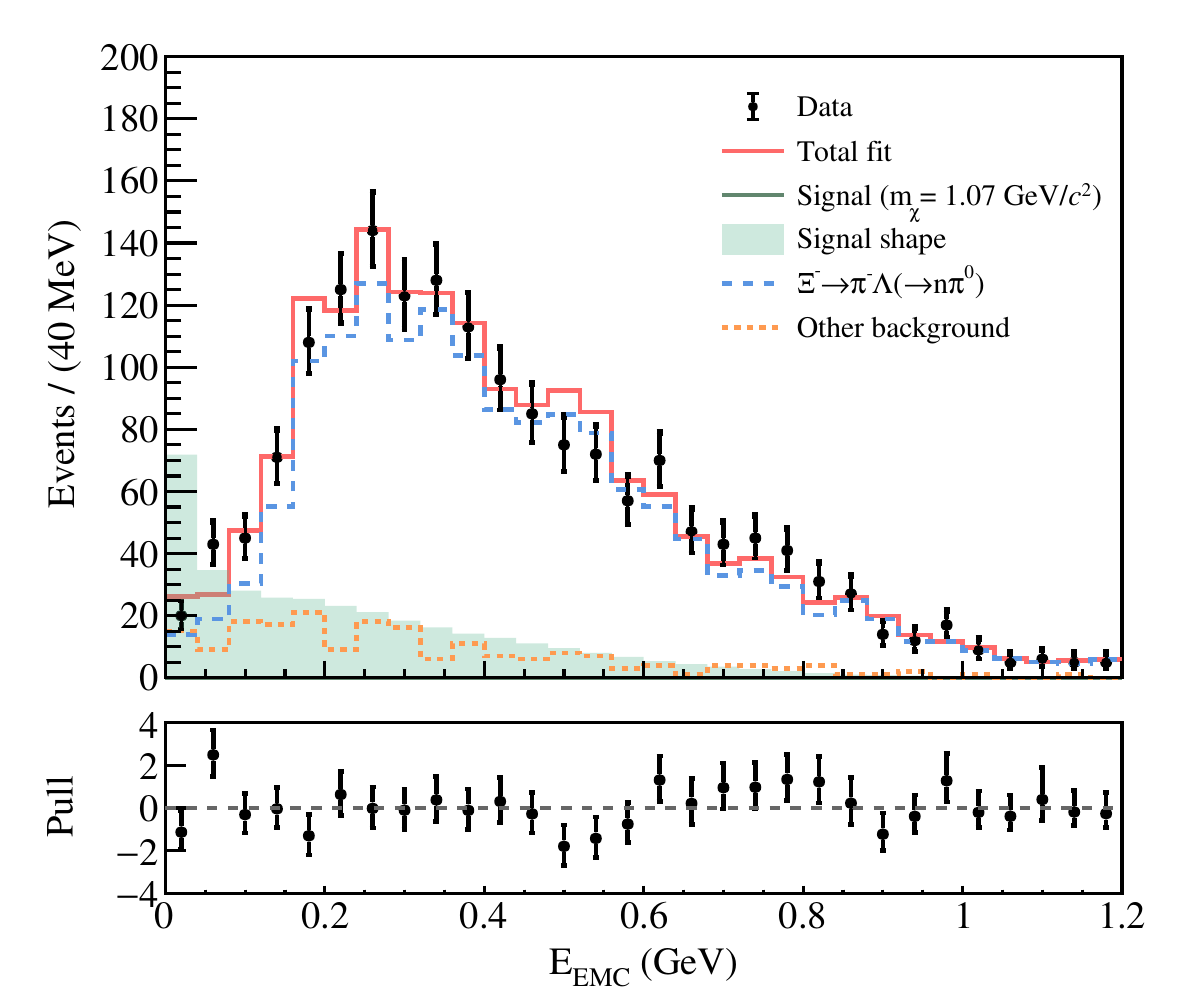}
  \put(20, 70){$(a)$}
  \end{overpic}
  \begin{overpic}[width=0.33\textwidth]{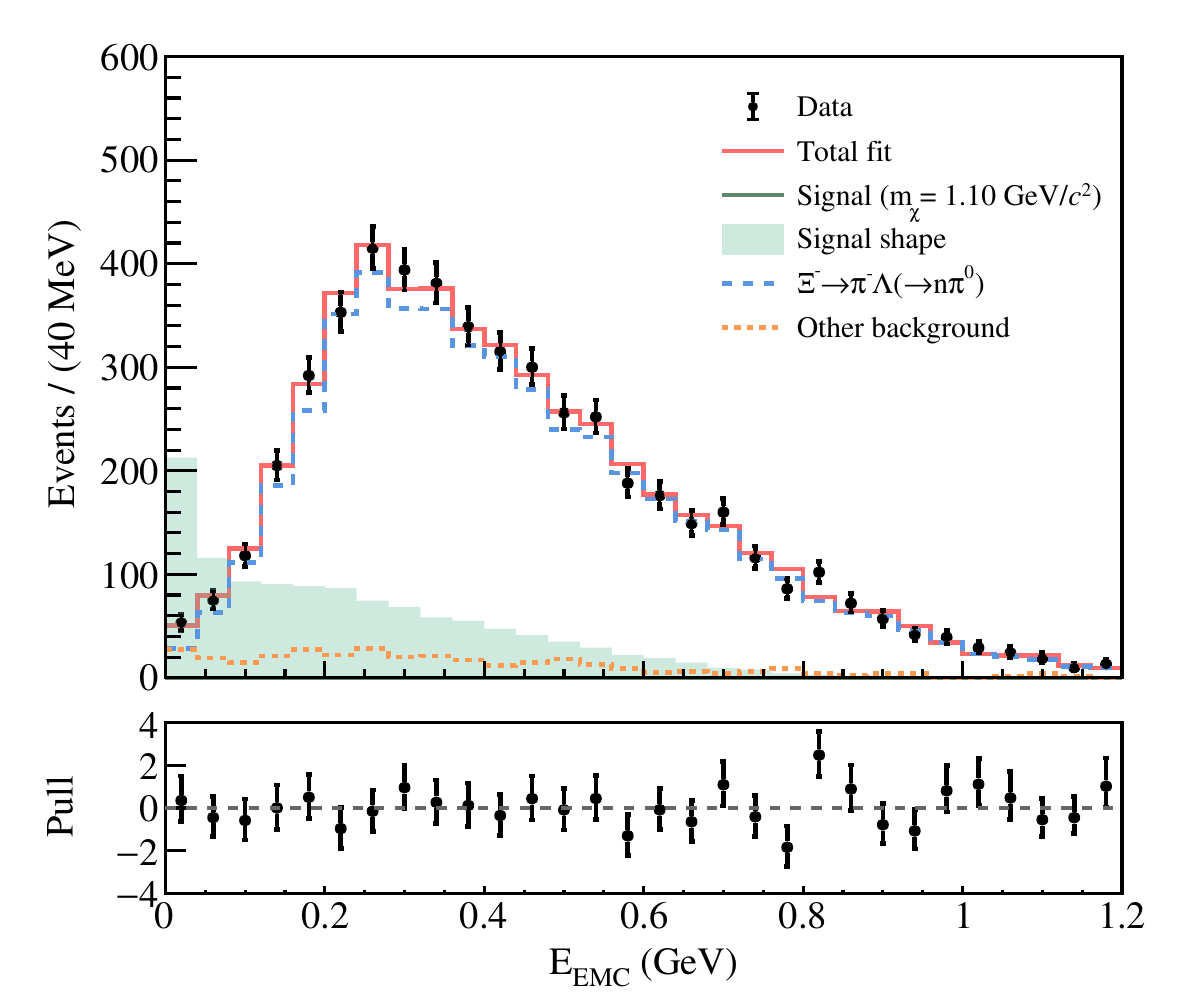}
  \put(20, 70){$(b)$}
  \end{overpic}
  }
\mbox
  {
    \begin{overpic}[width=0.33\textwidth]{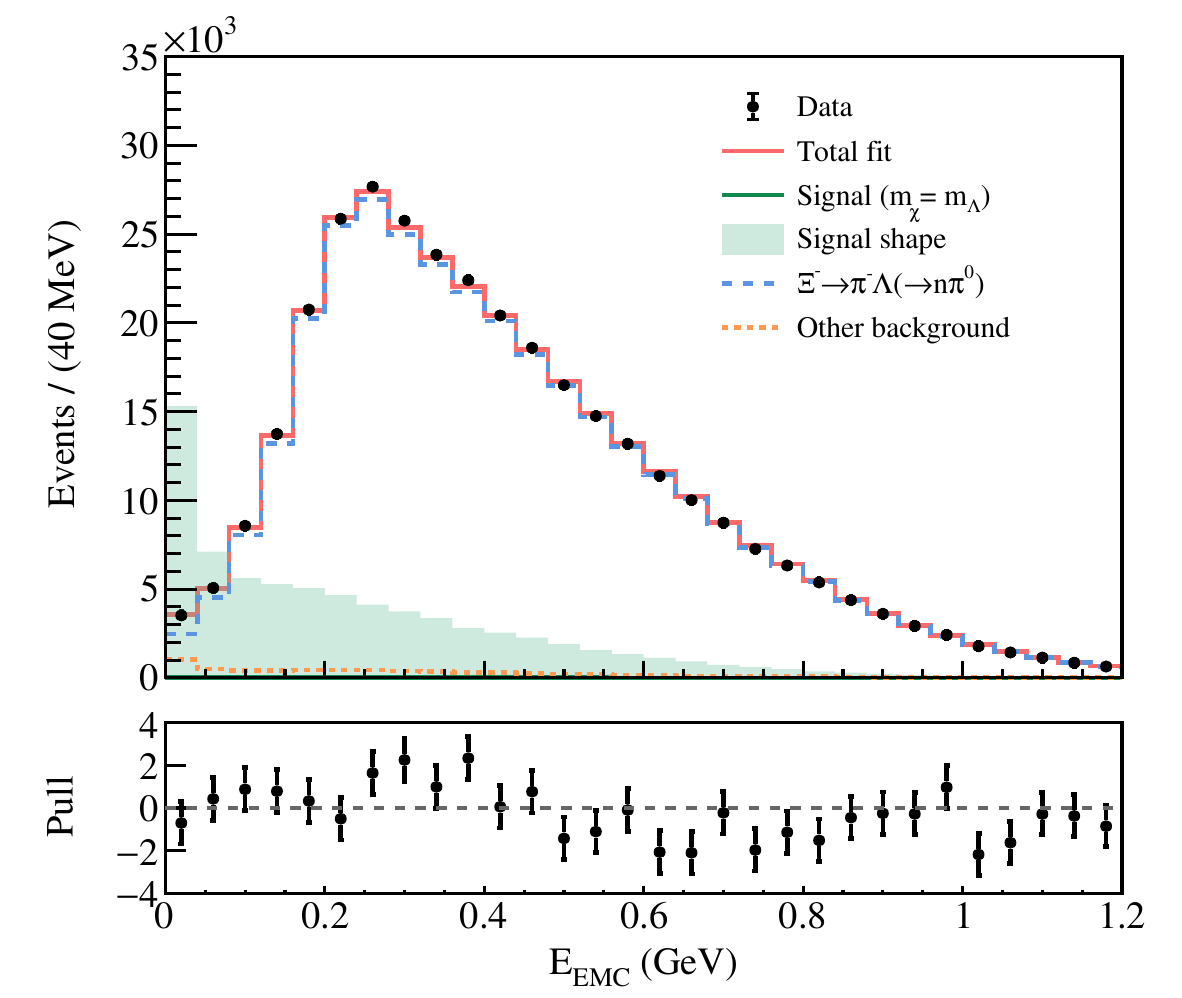}
  \put(20, 70){$(c)$}
  \end{overpic}
  \begin{overpic}[width=0.33\textwidth]{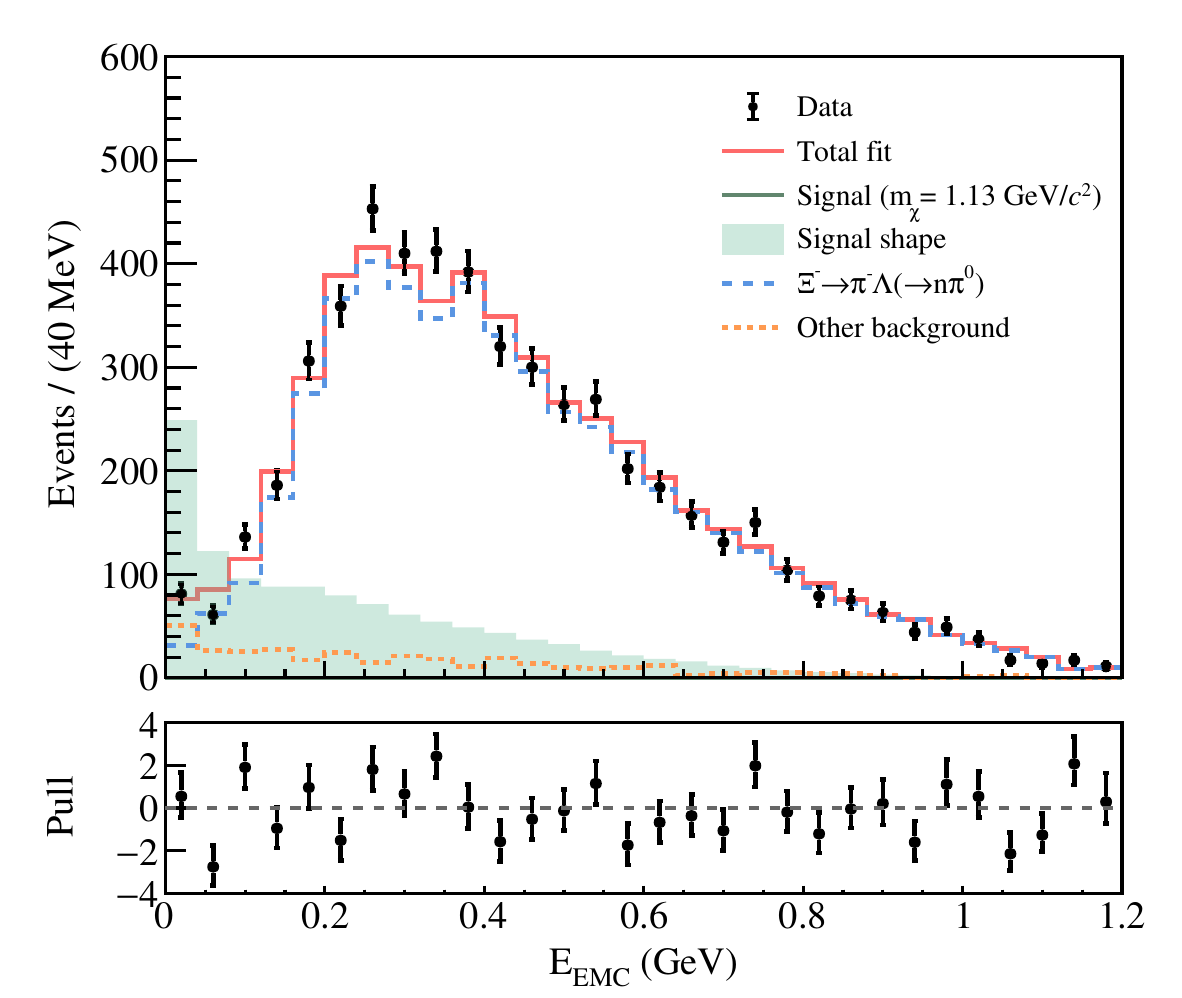}
  \put(20, 70){$(d)$}
  \end{overpic}
  \begin{overpic}[width=0.33\textwidth]{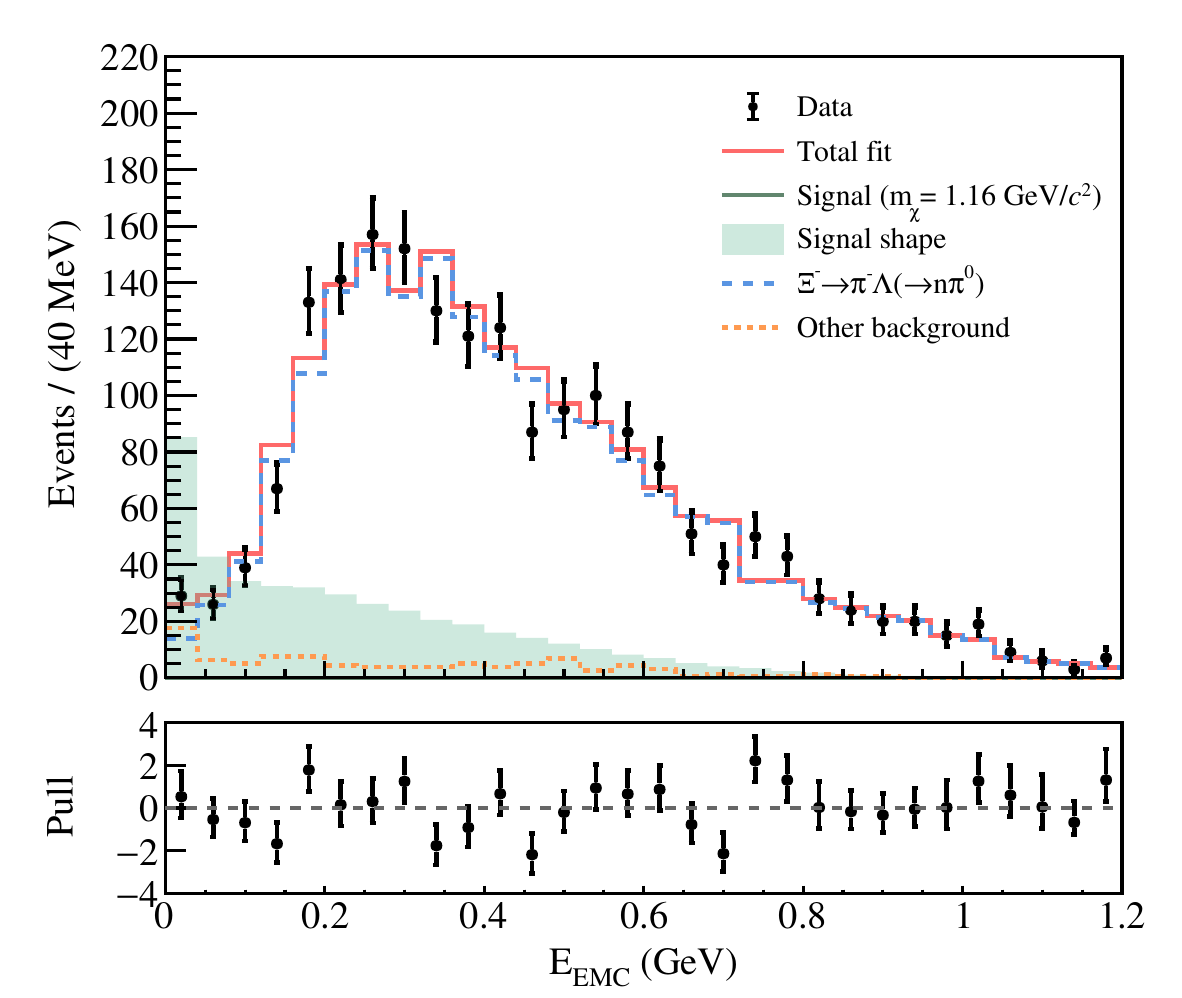}
  \put(20, 70){$(e)$}
  \end{overpic}
  }
 \caption{The fit results of the $\emc$ distributions under the dark baryon mass hypotheses of (a) $m_\chi=1.07\gevcc$, (b) $m_\chi=1.10\gevcc$, (c) $m_\chi=m_\lmd$, (d) $m_\chi=1.13\gevcc$, and (e) $m_\chi=1.16\gevcc$.
The black dots with error bars are data points and the red solid lines show the fit results. The signal, signal shape, $\xim\ra\pim\lmd(\ra n\piz)$ background, and other background in the inclusive MC sample are represented by the green solid line, green shaded region, blue dashed line, and orange dashed line, respectively.
The signal shape is normalized to BFs of (a) $5\times10^{-4}$, (b) $2\times10^{-3}$, (c) $1\times10^{-1}$, (d) $3\times10^{-3}$, and (e) $8\times10^{-4}$. The bottom panels show the fit residuals.
}
 \label{fig:fit_emc_drk}
\end{figure*}

The corrected distribution of $\emc$ is used as input in a binned maximum-likelihood fit to determine the DT signal yield. In the fit, the signal, background of $\xim\ra\pim\lmd(\ra n\piz)$, and other backgrounds in the inclusive MC sample are described by their MC-simulated shapes after the data-driven correction. The yields of signal and $\xim\ra\pim\lmd(\ra n\piz)$ background are free to float while the yield of other backgrounds in the inclusive MC sample is fixed.
The fit results of $\emc$ are shown in Fig.~\ref{fig:fit_emc_drk}. No significant signal is observed.
The DT yields of $\xim\ra\pim\inv$ are determined to be $(-13.3\pm23.9)$, $(-26.9\pm35.6)$, $(163.5\pm266.8)$, $(-26.9\pm37.5)$, and $(-28.3\pm22.7)$ under the $m_\chi$ hypotheses of 1.07$\gevcc$, 1.10$\gevcc$, $m_\lmd$, 1.13$\gevcc$, and 1.16$\gevcc$, respectively, where the uncertainties are only statistical.

\section{Systematic uncertainty}

By employing the DT technique in the analysis, the systematic uncertainties associated with the ST selection can be  canceled out. The remaining systematic uncertainties are divided into two types: additive and multiplicative. The additive uncertainties are related to the specific fit methods of DT, while the multiplicative uncertainties are associated with the knowledge of the signal efficiency and ST yields.

When performing the binned maximum-likelihood fit to the $\emc$ distribution, the uncertainty arising from the choice of bin width is considered by using alternative bin widths of 50 MeV and 30 MeV. The uncertainty due to the signal shape is assigned by considering alternative signal models in which the decay-asymmetry parameter of the $\xim\ra\pim\inv$ decay is varied between -1 and 1. The uncertainty due to the background shape of the $\xim\ra\pim\lmd(\ra n\piz)$ process is studied with the alternative background model by changing the decay parameter of the $\lmd\ra n\piz$ process by $\pm1\sigma$~\cite{Lmd_para}. 
The decay parameter of the $\lmd\ra p\pim$ process affects the $\pb$ momentum distribution at the ST side and is varied by $\pm1\sigma$~\cite{Xim_para} to obtain the alternative signal and $\xim\ra\pim\lmd(\ra n\piz)$ background shapes.
For other backgrounds in the inclusive MC sample, the fixed yield is shifted
within $\pm1\sigma$ of statistical uncertainty, and the alternative shape is obtained by smoothing the original shape with the kernel density estimation method~\cite{KDE}. The fit is performed twelve times in total with different methods, and the maximum upper limits (UL) are recorded.

The multiplicative systematic uncertainties are listed in Table~\ref{tab:multi}. The uncertainty due to the ST yield is evaluated by replacing the background shape from a 2nd-order Chebyshev polynomial to a 3rd-order and a 1st-order Chebyshev polynomial function. The uncertainties arising from pion tracking and PID, $\chi_{\pim\chi}^2$, $\chi_{\pim\lmd}^2>\chi_{\pim\chi}^2$, and $P_{\pim}$ requirements are assigned from studies of a control sample of $\xim\ra\pim\lmd(\ra n\piz)$ decays, where the efficiency difference between data and MC simulation is taken as the uncertainty.
The uncertainty of the signal model is obtained from signal MC samples with different decay parameters. By assuming all the sources to be independent, the total multiplicative systematic uncertainties are determined to be 5.1\%, 8.3\%, 2.9\%, 5.3\%, and 4.6\% under the $m_\chi$ hypotheses of 1.07$\gevcc$, 1.10$\gevcc$, $m_\lmd$, 1.13$\gevcc$, and 1.16$\gevcc$, respectively.

\begin{table*}[htbp]
\centering
\caption{Summary of multiplicative systematic uncertainties for different $m_\chi$ hypotheses.}
\label{tab:multi}
\begin{tabular}{c|ccccc}
\hline \hline Source & \multicolumn{5}{c}{Uncertainty (\%)}  \\
\hline
$m_\chi$ & $1.07\gevcc$ &$1.10\gevcc$ &$m_\lmd$ & $1.13\gevcc$ &$1.16\gevcc$\\
\hline
ST yield & 0.2 & 0.2 & 0.2 &0.2 &0.2 \\
Tracking and PID& 1.4 & 1.4 & 1.4 &1.4 &1.4\\
$\chi^2_{\pim\chi}$ requirement & 0.1& 0.1 & 0.1 &0.1 &0.1\\
$\chi^2_{\pim\lmd}>\chi^2_{\pim\chi}$  & 0.1& 5.0 & - &4.0 &0.1\\
$P_{\pim}$ requirement& 4.7& 6.3 & 1.3 &0.2 &3.3\\
Signal model &1.3 &1.3 &2.2 &3.2 &2.9 \\
\hline  Total (multiplicative) & 5.1  & 8.3 & 2.9& 5.3& 4.6\\
\hline
\end{tabular}
\end{table*}

\section{Result} 

\begin{figure}[H]
  \begin{center}
    \begin{overpic}[width=7.5cm]{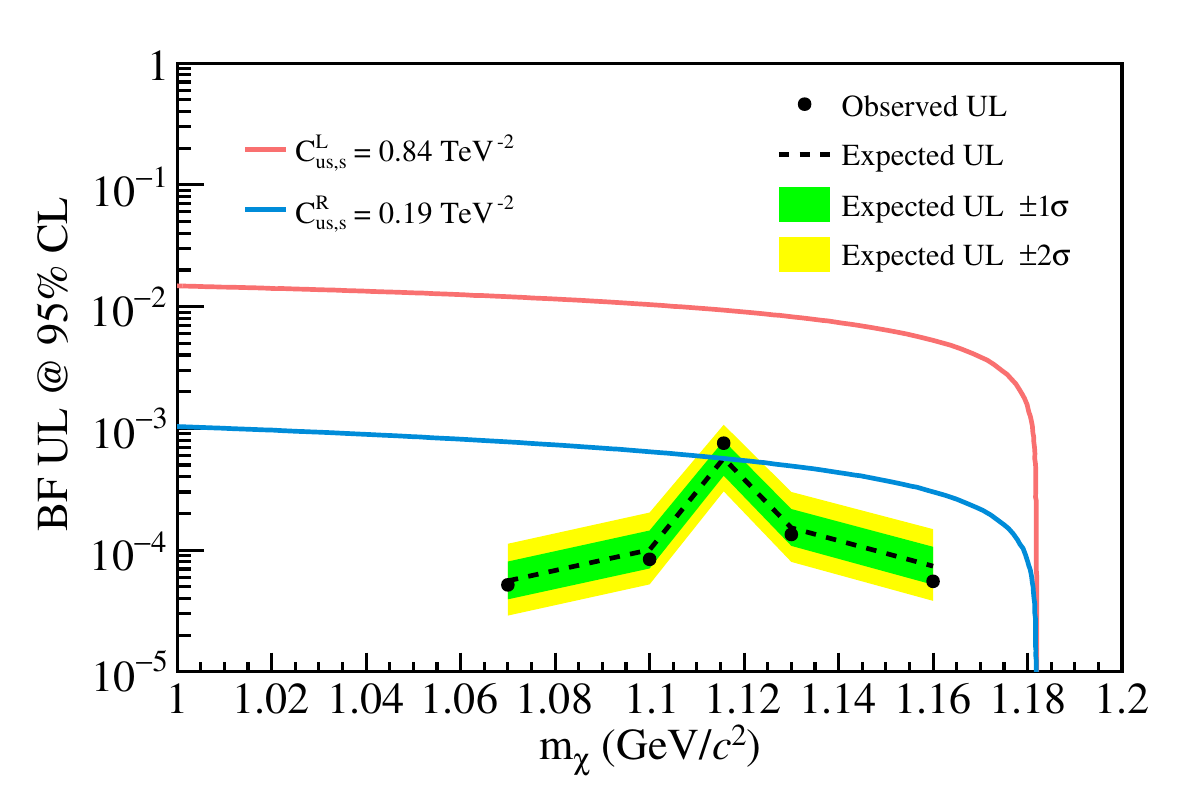}
    \end{overpic} 
    \caption{The expected and observed ULs at the 95\% C.L. on the BF of $\xim\ra\pim\inv$ under different $m_\chi$ hypotheses. The blue and red lines represent the maximum allowed BFs, where the Wilson coefficients are set to $C_{u s, s}^L<0.84\, \mathrm{TeV}^{-2}$ and $C_{u s, s}^R<0.19\, \mathrm{TeV}^{-2}$ derived from the LHC searches for the colored mediators~\cite{dark_baryons}.}
    \label{fig:couple}
  \end{center}
\end{figure}

\begin{figure}[H]
  \begin{center}
    \begin{overpic}[width=7.5cm]{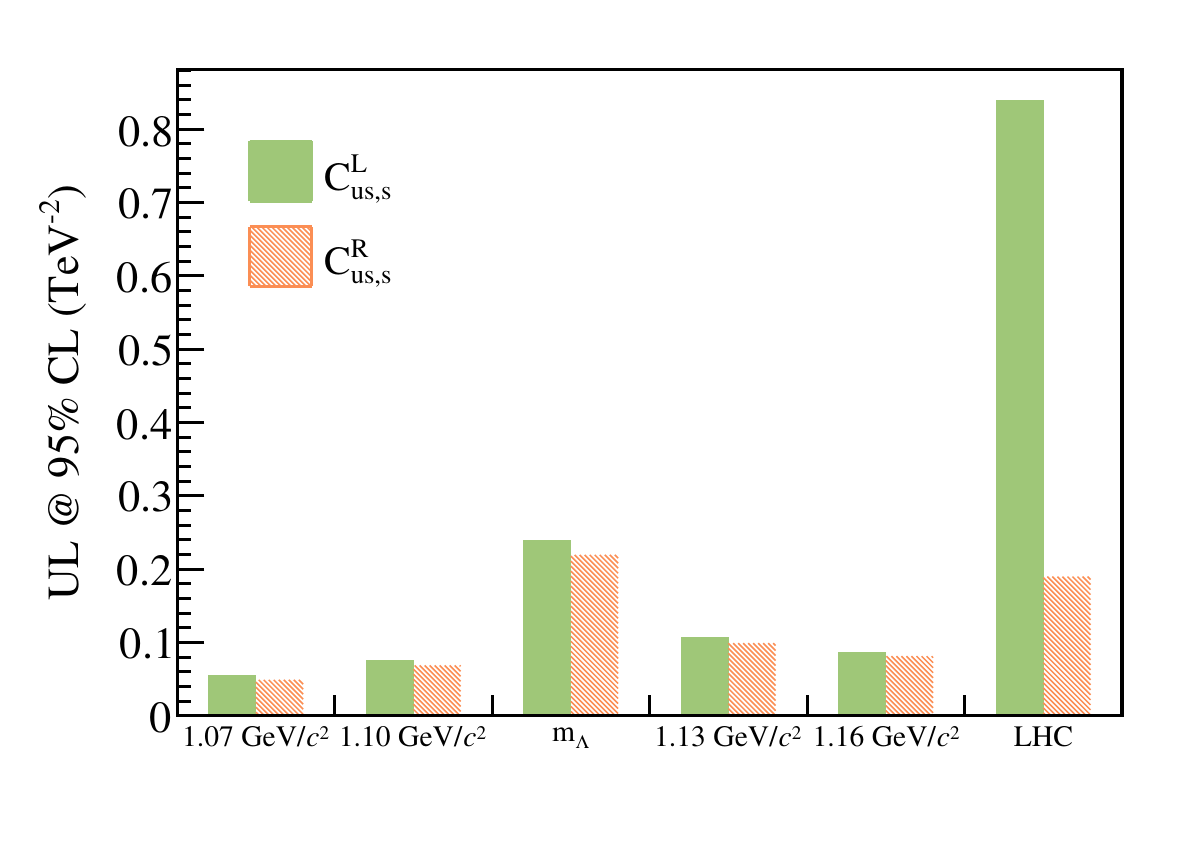}
    \end{overpic} 
    \caption{The 95\% C.L. ULs on the Wilson coefficients $C_{u s, s}^L$ and $C_{u s, s}^R$ derived from the results under different $m_\chi$ hypotheses. The constraints derived from the LHC searches~\cite{dark_baryons} are also shown.}
    \label{fig:wilson}
  \end{center}
\end{figure}

Since no significant signal is observed in the data samples, a one-sided frequentist profile-likelihood method \cite{CLs} is used to compute the expected and observed ULs on $\mathcal{B}(\xim\ra\pim\inv)$, where the total multiplicative systematic uncertainty is included in the overall likelihood as a Gaussian nuisance parameter with a width equal to the uncertainty. The 90\% (95\%) C.L. ULs are determined to be $4.2\times10^{-5}$ ($5.2\times10^{-5}$), $6.9\times10^{-5}$ ($8.4\times10^{-5}$), $6.5\times10^{-4}$ ($7.6\times10^{-4}$), $1.1\times10^{-4}$ ($1.3\times10^{-4}$), and $4.5\times10^{-5}$ ($5.5\times10^{-5}$), under the $m_\chi$ hypotheses of 1.07$\gevcc$, 1.10$\gevcc$, $m_\lmd$, 1.13$\gevcc$, and 1.16$\gevcc$, respectively. The 95\% C.L. expected and observed ULs are shown in Fig.~\ref{fig:couple}. 
The right-handed and left-handed effective operators $\mathcal{O}_{u s, s}^R$ and $\mathcal{O}_{u s, s}^L$ mediate the decay of $\xim$ into $\pim$ and the dark baryon $\chi$~\cite{dark_baryons}, with the corresponding Wilson coefficients denoted as $C_{u s, s}^L$ and $C_{u s, s}^R$, respectively. The BF of $\xim\ra\pim\chi$ decay is proportional to the square of the Wilson coefficients.
The 95\% C.L. ULs on the Wilson coefficients, derived from the results under various $m_\chi$ hypotheses, are shown in Fig.~\ref{fig:wilson}. The result under the $m_\chi=1.07\gevcc$ hypothesis corresponds to the 95\% C.L. constraints of $C_{u s, s}^L<5.5\times10^{-2}\, \mathrm{TeV}^{-2}$ and $C_{u s, s}^R<4.9\times10^{-2}\, \mathrm{TeV}^{-2}$. The constraints obtained on the Wilson coefficients $C_{u s, s}^L$ and $C_{u s, s}^R$ significantly improve previous limits derived from the LHC searches for the colored mediators~\cite{dark_baryons} by factors of 15 and 4, respectively.

\section{Summary}
A search for a dark baryon in the two-body decay $\xim\ra \pim\inv$ is presented using $(10.087\pm0.044)\times10^{9}$ $\jpsi$ events collected at a CM energy of $\sqrt{s}=3.097\gev$ with the BESIII detector at the BEPCII collider. No significant signal is observed and the 90\% (95\%) C.L. ULs on $\mathcal{B}(\xim\ra\pim\inv)$ are determined to be $4.2\times10^{-5}$ ($5.2\times10^{-5}$), $6.9\times10^{-5}$ ($8.4\times10^{-5}$), $6.5\times10^{-4}$ ($7.6\times10^{-4}$), $1.1\times10^{-4}$ ($1.3\times10^{-4}$), and $4.5\times10^{-5}$ ($5.5\times10^{-5}$), under the dark baryon mass hypotheses of 1.07$\gevcc$, 1.10$\gevcc$, $m_\lmd$, 1.13$\gevcc$, and 1.16$\gevcc$, respectively.  The constraints obtained on the Wilson coefficients $C_{u s, s}^L$ and $C_{u s, s}^R$ are more stringent than the previous limits from the LHC searches for the colored mediators.

\section*{Acknowledgement}

The BESIII Collaboration thanks the staff of BEPCII (https://cstr.cn/31109.02.BEPC) and the IHEP computing center for their strong support. This work is supported in part by National Key R\&D Program of China under Contracts Nos. 2023YFA1606000, 2023YFA1606704, 2020YFA0406400; National Natural Science Foundation of China (NSFC) under Contracts Nos. 11635010, 11935015, 11935016, 11935018, 12025502, 12035009, 12035013, 12061131003, 12192260, 12192261, 12192262, 12192263, 12192264, 12192265, 12221005, 12225509, 12235017, 12361141819; the Chinese Academy of Sciences (CAS) Large-Scale Scientific Facility Program; CAS under Contract No. YSBR-101; 100 Talents Program of CAS; The Institute of Nuclear and Particle Physics (INPAC) and Shanghai Key Laboratory for Particle Physics and Cosmology; Agencia Nacional de Investigación y Desarrollo de Chile (ANID), Chile under Contract No. ANID PIA/APOYO AFB230003; German Research Foundation DFG under Contract No. FOR5327; Istituto Nazionale di Fisica Nucleare, Italy; Knut and Alice Wallenberg Foundation under Contracts Nos. 2021.0174, 2021.0299; Ministry of Development of Turkey under Contract No. DPT2006K-120470; National Research Foundation of Korea under Contract No. NRF-2022R1A2C1092335; National Science and Technology fund of Mongolia; Polish National Science Centre under Contract No. 2024/53/B/ST2/00975; Swedish Research Council under Contract No. 2019.04595; U. S. Department of Energy under Contract No. DE-FG02-05ER41374

\end{multicols}

\end{document}